\documentclass[sigconf]{acmart}

\usepackage{enumitem}
\usepackage{afterpage}
\usepackage{listings}
\usepackage{xcolor}
\usepackage{graphicx}
\usepackage{float}
\usepackage{caption}
\usepackage{subcaption}
\usepackage{placeins}
\usepackage{url}
\usepackage{epstopdf}

\usepackage{booktabs}
\usepackage{footnote}
\makesavenoteenv{table}
\makesavenoteenv{tabular}

\usepackage{paralist}

\begin{document}

\setlength{\abovedisplayskip}{2pt}
\setlength{\belowdisplayskip}{2pt}
\setlength{\belowcaptionskip}{-0.4cm}

\title{ns-3 Implementation of the 3GPP MIMO\\Channel Model for Frequency Spectrum above 6 GHz}

\author{\texorpdfstring{Menglei Zhang$^\diamond$, Michele Polese$^*$, Marco Mezzavilla$^\diamond$, Sundeep Rangan$^\diamond$, Michele Zorzi$^*$\\
\small $^\diamond$NYU Tandon School of Engineering, New York University, Brooklyn NY, USA 
\\\small e-mail: \{menglei, mezzavilla, srangan\}@nyu.edu \\
\small $^*$Department of Information Engineering, University of Padova, Italy \\
 \small e-mail: \{polesemi, zorzi\}@dei.unipd.it}{}}

\setcopyright{none}
\settopmatter{printacmref=false, printccs=true, printfolios=true}

\begin{abstract}
Communications at mmWave frequencies will be a key enabler of the next generation of cellular networks, due to the multi-Gbps rate that can be achieved. However, there are still several problems that must be solved before this technology can be widely adopted, primarily associated with the interplay between the variability of mmWave links and the complexity of mobile networks. An end-to-end network simulator represents a great tool to assess the performance of any proposed solution to meet the stringent 5G requirements. Given the criticality of channel propagation characteristics at higher frequencies, we present our implementation of the 3GPP channel model for the 6-100 GHz band for the ns--3 end-to-end 5G mmWave module, and detail its associated MIMO beamforming architecture.
\end{abstract}



\maketitle
\vspace{-1mm}
\section{Introduction}
Thanks to the wide available spectrum at higher frequencies, mmWave communications will likely play a major role in the future of cellular networks. The 3GPP is considering the usage of mmWaves for 5G New Radio (NR), the next generation of mobile networks. However, several issues must be solved before this technology becomes market-ready, such as sensitivity to sudden blockage, link outages and service unavailability, mobility management and interaction with legacy networks and protocols. In~\cite{ford2016framework} we presented an ns--3 module for the simulation of mmWave end-to-end networks, which has been used to study, among others, the performance of TCP over mmWave links~\cite{zhang2016transport} and of LTE-mmWave multi connectivity architectures~\cite{polese2016improved}. The channel model was limited to the measurements and statistics in~\cite{akdeniz2014millimeter}, which supported only the 28 and 73 GHz frequencies and did not provide consistency in space for mobility-based simulations. 

\textbf{Contribution 1:} In this paper, we present the implementation for the ns--3 mmWave module of the 3GPP channel model, described in~\cite{38900}. According to the 3GPP, this model should be used for the evaluation of NR technologies that rely on communications in the 6-100 GHz band, hence making it an extremely important feature for any mmWave system level simulator. Moreover, we combined the model with the ns--3 buildings module in order to simulate complex and realistic 3D scenarios, and implemented two of the optional features of the model, the \emph{spatial consistency} and the \emph{blockage model}, in order to be as compliant as possible with~\cite{38900}. 

\textbf{Contribution 2:} The second contribution of this paper is the MIMO beamforming architecture, which models two beamforming computational methods, and can be easily customized to study, for example, beam tracking solutions, beamforming vector calculation techniques, and different antenna architectures.

The paper is organized as follows. In Sec.~\ref{sec:soa}, we present the state of the art in mmWave channel modeling. The 3GPP channel model is described in Sec.~\ref{sec:3gppchannel}. Sec.~\ref{sec:channelImpl} reports a detailed description of the implementation of the basic channel model, and provides information on the simulator architecture and on the classes involved in the propagation computation and fading generation. Sec.~\ref{sec:bf} describes the MIMO beamforming architecture, and Sec.~\ref{sec:opt} illustrates the implementation of the additional channel features, along with some simulation results. Finally, in Sec.~\ref{sec:conclusions} we draw some conclusions and share our some possible directions for future research.

\vspace{-1.5mm}
\section{MmWave Channel Models}
\label{sec:soa}
In recent years the industrial and academic communities have developed several channel models for mmWave communications. Many of the proposed models extend the 3D-geometry based approach first proposed in the 3GPP Spatial Channel Model (SCM)~\cite{25996}, with the extensions of WINNER and WINNER-II~\cite{winnerII} and the adaptation to LTE systems~\cite{36873}.
For example, the METIS model~\cite{metisChannel} covers a frequency range from 380 MHz to 86 GHz, and provides different modeling methodologies: map-based, which relies on ray-tracing in a 3D environment, stochastic, that extends the SCM model, and hybrid, which combines the two. They address different needs in terms of a tradeoff between the complexity and the level of detail. Also the ITU-R M channel model~\cite{itum} is based on SCM, and extends it by modeling also of atmospheric loss, antenna array and semiconductor technologies for mmWave communications, in urban dense environments. The Quasi Deterministic Radio Channel Generator (QuaDRiGa) model, from the Fraunhofer Institute~\cite{jaeckel2014quadriga}, adds the support for consistent user mobility, massive MIMO at several frequencies (10, 28, 43, 60, 82 GHz).

A different approach is adopted by the European project MiWEBA, which proposed a quasi-deterministic channel model~\cite{miweba} for the outdoor propagation at 60 GHz, with the combination of a strong component modeled applying Friis' Law to the geometry of the scenario and few random components. Another channel model for the 60 GHz band is the 802.11ad model~\cite{jacob2011ray}, which uses ray-tracing and focuses on indoor propagation.
The COST 2100~\cite{liu2012cost} model deploys obstacles (scattering objects) in the simulation scenario, and analyzes which multipath components can reach a specific mobile terminal in a certain time and frequency interval. 

Finally, the NYU channel model based on measurements in New York City and described in~\cite{akdeniz2014millimeter} supports the 28 and 73 GHz carrier frequencies, and is already implemented in the NYU mmWave module for ns--3~\cite{ford2016framework}.

\begin{table*}[t]
\renewcommand{\arraystretch}{1}
\small
\setlength{\belowcaptionskip}{0.1cm}
\setlength{\abovecaptionskip}{-0.3cm}
\centering
\begin{tabular}{@{}lllll@{}}
\toprule
								& UMi \hspace{1cm} 	& UMa 	& RMa 		& In  \\ \midrule
Carrier frequency $f_c$ [GHz] & 6-100 & 6-100 & 6-7 & 6-100 \\

ISD [m] 						& 200	& 500	& 1732-5000	& 20  \\
BS antenna height $h_{BS}$ [m] 	& 10	& 25	& 35		& 3  \\
UT locations					& \multicolumn{2}{l}{80\% indoor, 20\% outdoor} & 50\% in cars, 50\% outdoor & Indoor\\
UT height $h_{UT}$ [m] 			&  [1.5, 22.5] & [1.5, 22.5] & 1.5 & 1 \\
UT mobility $v$ (horizontal plane) [km/h] &3 & 3 & N/A & 3 \\
Minimum BS-UT distance$^*$ (2D) $d_{2D, \min}$ [m] & 10 & 35 & 35 & 0 \\
UT distribution 				& \multicolumn{4}{c}{Uniform} \\
\bottomrule
\end{tabular}
\caption{Main parameters for the different scenarios.\\ \footnotesize $^*$The pathloss equations in~\cite{38900} allow to reach a $d_{2D, \min} = 10$ m for UMi, UMa and RMa.}
\label{tab:scenarioparams}
\end{table*}

\section{3GPP TR 38.900 Channel Model}
\label{sec:3gppchannel}

The 3GPP Technical Report 38.900~\cite{38900} contains a detailed channel model for the frequency band from 6 to 100 GHz. 
It is applicable for bandwidth up to 10\% of the carrier frequency $f_0$ (with a limit at 2 GHz), and accounts for the mobility of one of the two terminals (i.e., in typical cellular networks the base station is fixed and the user terminal moves). Finally, it provides several optional features that can be plugged in to the basic model, in order to simulate spatial consistency (i.e., the radio environment conditions of close-by users are correlated), blockage and oxygen absorption.
This model supports different scenarios, which must be chosen when setting the simulations parameters: 
\begin{compactitem}
	\item Urban Microcell (UMi), with a street canyon assumption, i.e., the BSs are below the rooftop of the buildings surrounding the street, with small Inter-Site Distance (ISD) between base stations, as for example in a densely deployed urban area. The scenario considers an outdoor BS with both indoor and outdoor users, i.e., it is possible to model Outdoor-to-Indoor (O2I) and Outdoor-to-Outdoor (O2O) propagation;
	\item Urban Macrocell (UMa), with BSs above the roofs of buildings with larger ISD than in the UMi scenario.
	\item Rural Macrocell (RMa), which focuses on a rural deployment with a larger coverage area than the urban case, with the ISD in the order of thousands of meters. Currently, this scenario is validated only for frequencies up to 7 GHz.
	\item Indoor Office (In), that captures the dynamics of offices, with cubicles and BSs on ceilings, and shopping malls, which are several stories high and with an open area shared by different floors. The scenarios are defined as either open (InOO) or mixed (InMO), but the definitions are inconsistent throughout the 3GPP report, since in certain parameter tables the scenarios are shopping mall and indoor office, while in other tables there is just a single indoor scenario. We chose to support only the InOO and InMO scenarios in the whole framework, since their parameters are available both for the pathloss and the fading computations.
\end{compactitem}
The main parameters that characterize each scenario are reported in Table~\ref{tab:scenarioparams}. These scenarios cover a wide range of 5G possible deployments and use cases: ultra-dense urban area, with micro and macro base stations, wireless backhaul with BSs over rooftop or on street lamp posts, device-to-device and vehicle-to-vehicle communications. Moreover, thanks to the RMa scenario, it is possible to investigate if mmWave systems can be used with current network deployments.

\textbf{Channel Model: }
%
%
The channel model is a 3D statistical spatial channel model whose design is an evolution of the 3GPP SCM model. It is based on measurement campaigns of propagation at mmWave frequencies conducted by several 3GPP partners. 
Given a certain scenario, the 3D positions, velocities and indoor/outdoor state of User Terminals (UTs) and Base Stations (BSs), the model assigns a LOS/NLOS condition to each link according to scenario-specific probability distributions. This condition affects both the pathloss computation and the generation of fading parameters. For the pathloss, different formulas are provided, according to the scenario, and an additional O2I component can be added, with the possibility of modeling different kind of building penetration losses (glass, concrete, wood). 

The channel is described by a channel matrix $\mathbf{H}(t, \tau)$, of size $U\times S$, where $U$ and $S$ are the number of antennas at the receiver and the transmitter. Each entry depends on $N$ different multipath components, the \textit{clusters}, which have different delay and received power, according to an exponential power delay profile. The clusters are themselves a combination of $M$ \textit{rays}, each with a slightly different arrival and departure angle in the vertical and horizontal planes. All these parameters are randomly drawn from distributions specified in the technical report, which vary according the scenario, the LOS condition, the indoor or outdoor state, the frequency, the distance between transmitter and receiver and their heights. It is possible to distinguish two different fading components. The large scale fading depends on user mobility or changes in the scenario, that have an impact on parameters such as the delay spread (DS), the angular spread of arrival and departure, either in the azimuth, i.e., the horizontal plane (ASA, ASD), or in the zenith, i.e., the vertical plane (ZSA, ZSD), the Ricean factor for the LOS condition (K) and the shadow fading. The large scale parameters for each link are generated and cross-correlated using the procedure originally suggested in the WINNER-II channel model~\cite{winnerII}. Different correlation parameters are given for the scenarios previously introduced and for the LOS and NLOS conditions. The small scale fading, instead, depends on the small variation of the various multipath components, i.e., the delay of the different clusters, the arrival and departure angles of the rays, the variation in the power of the different clusters and the Doppler effect introduced by a moving UT. The main difference between the LOS and the NLOS conditions in the channel matrix $\mathbf{H}(t, \tau)$ is the presence of a Ricean component, i.e., of a cluster with a much higher power than the others.

\textbf{Optional Features: }
The 3GPP TR 38.900 specifies additional features that can be plugged into the flow described in the previous paragraphs in order to increase the level of detail of the channel model, if the simulation requires it. For example, for drop-based simulations with no mobility the basic channel model is enough, but, if the UT moves, more realistic results may be obtained with a channel model that is updated correlating the $\mathbf{H}(t, \tau)$ matrix in the new position with that in the previous one. This is obtained using the \textit{spatial consistency} feature, that provides the decorrelation distances for the fading parameters, the LOS/NLOS condition and the indoor/outdoor state. Another optional feature is the \textit{oxygen absorption} one, that introduces a frequency-dependent additional pathloss factor to account for the atmospheric absorption at certain frequencies. The \textit{blockage} option allows to attenuate the receive power of certain clusters according to randomly deployed blockers: this simulates the presence of human and vehicular obstacles. 

\section{ns--3 Implementation}
\label{sec:channelImpl}
The ns--3 modules that handle the wireless channel are the Propagation module, which defines the \texttt{PropagationLossModel} interface, and the Spectrum module, introduced in~\cite{baldo2009spectrum}, with the \texttt{Spectrum\-PropagationLossModel} interface. By extending the first, it is possible to implement different propagation models, while the second is the basis for modeling small scale fading in terms of a Power Spectral Density (PSD). In the mmWave module, the physical layer at the BS or the UT side uses \texttt{MmWaveSpectrumPhy} to simulate actual transmissions with noise and interference. It is this class which is in charge of computing the SINR of the transmissions, given the propagation conditions specified in the mmWave module extensions of \texttt{PropagationLossModel} and \texttt{SpectrumPropagationLossModel}. In \cite{mezzavilla20155g,ford2016framework} we described the other channel models available in the mmWave module: a ray-tracing model implemented in the class \texttt{MmWaveRayTracing} and the NYU statistical model~\cite{akdeniz2014millimeter} implemented in \texttt{MmWavePropagationLossModel} and \texttt{MmWaveBeamforming}.

\begin{figure}[!t]
\includegraphics[width=0.8\columnwidth,trim = 1.8cm 3.5cm 0cm 2cm,clip] {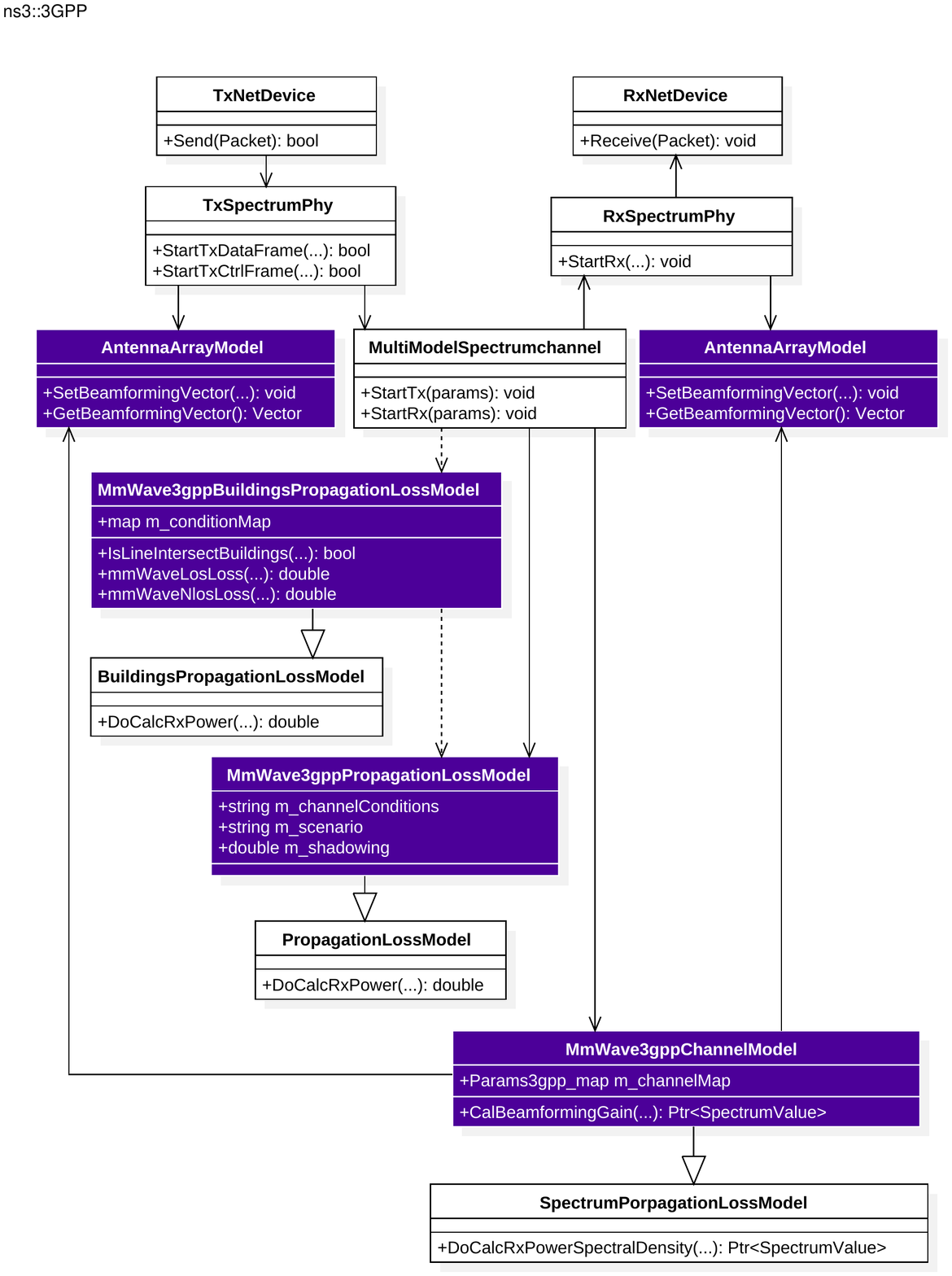}
\caption{Class UML diagram, the classes related to the 3GPP channel model are marked in purple.}
\label{fig:uml}
\end{figure}

In this paper we introduce the implementation for ns--3 of the 3GPP channel model described in Sec.~\ref{sec:3gppchannel}. The propagation and shadowing are computed in the \texttt{MmWave3gppPropagationLossModel}, or, if the ns--3 buildings module is used, in the \texttt{MmWave3gppBuilding\-PropagationLossModel}. The channel matrix $\mathbf{H}(t, \tau)$ is computed in the class \texttt{MmWave3gppChannel}, which computes also the beamforming vector and interacts with the \texttt{AntennaArrayModel} class. A basic UML diagram that describes the relationships among these classes is shown in Fig.~\ref{fig:uml}, and the source code can be found in the NYU mmWave ns--3 repository\footnote{\url{https://github.com/nyuwireless/ns3-mmwave}}.

\textbf{Tutorial Example:} We also provide an ns--3 script that can be used to run a simulation using the 3GPP channel model in the \texttt{MmWave3gppChannelExample} example, which models the scenario shown in Fig.~\ref{fig:building} with two BSs and two UTs. When designing its own script, the researcher can select one of the two propagation models by setting the \texttt{PathlossModel} attribute of the \texttt{MmWaveHelper} object that is used to set up the simulation, and set the \texttt{ChannelModel} attribute to \texttt{"ns3::MmWave3gppChannel"}. Then the other parameters that must be chosen are the scenario, the LOS condition model, the usage of the optional NLOS pathloss computations, spatial consistency and blockage features. All the attributes related to this channel model can be found in the provided example. 

Our implementation supports all the scenarios described in Sec.~\ref{sec:3gppchannel}, with outdoor only and outdoor to indoor propagation. Moreover, besides the basic channel model, we support two optional features: spatial consistency and blockage.
In the following paragraphs we will describe in detail how the previously mentioned classes are related to the 3GPP channel model and which are the implementation assumptions we made.

\subsection{Antenna Modeling}
The antenna is modeled in the class \texttt{AntennaArrayModel} as a Uniform Planar Array (UPA), i.e., as a rectangular panel. In the current implementation we support a single rectangular panel per BS/UT, with $C$ columns and $R$ rows. We plan to extend the model in the future in order to account for a multi-panel structure, and multiple sectors per BS site as required by~\cite{38900}. 

\label{sec:prop}
\begin{figure*}[t]
	\centering
	\includegraphics[width= 0.75\textwidth,trim = 6cm 0cm 16cm 0.5cm,clip ]{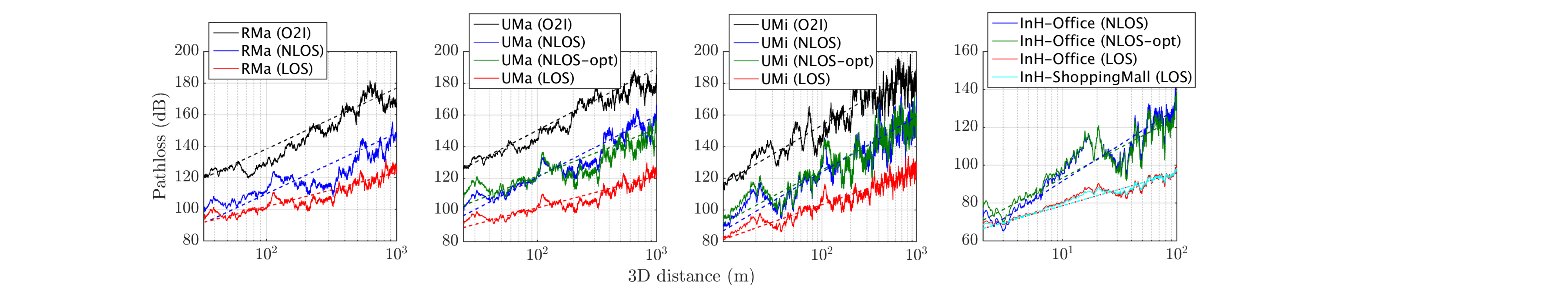}
	\caption{Pathloss with and without (dashed line) shadowing for the 3 outdoor scenarios, also with O2I penetration loss, and the two indoor scenarios.}
	\label{fig:pathloss}
\end{figure*}

The number of antenna elements for each device can be set using the \texttt{AntennaNum} attribute of the \texttt{MmWaveEnbNetDevice} and \texttt{MmWaveUeNetDevice} classes. This allows to specify a different number of antennas for the BSs and the UTs, as it would happen in a realistic deployment. The default values are 64 antennas for the BS, arranged in a square with $8\times8$ elements, and 16 for the UT, in a square with $4\times4$ elements. The horizontal and vertical spacing $d_H$ and $d_V$ are stored in two private variables of the \texttt{AntennaArrayModel} class, and expressed as multiples of the wavelength $\lambda = c/f_c$. The default value is 0.5, but it is possible to update it using the \texttt{AntennaHorizontalSpacing} and \texttt{AntennaVerticalSpacing} attributes. Given this information, the \texttt{AntennaArrayModel} class provides the method \texttt{GetAntennaLocation} which computes the coordinates of an antenna element, relative to the BS/UT position, i.e., the left bottom corner is in (0,0,0) and the $i$-th element is in position $(0, d_H(i\mod C), d_V \lfloor i/C \rfloor)$. Finally, the method \texttt{GetRadiationPattern} returns the antenna radiation pattern, given the angle of departure or arrival. In order to test different radiation patterns it is possible to implement them in this method.

\subsection{Propagation Loss Modeling}

The propagation loss, the LOS probability and the outdoor to indoor penetration models are described in~\cite[Sec.~7.4]{38900}. In our implementation, once one of the scenarios introduced in Sec.~\ref{sec:3gppchannel} is chosen, it is possible to select the LOS condition and the indoor/outdoor state statistically, geometrically or deterministically. The pathloss then depends on the UT status, the distance on the horizontal plane (2D distance) and the 3D distance from the UT to the BS. 

The class \texttt{MmWave3gppPropagationLossModel}, which extends the \texttt{PropagationLossModel} interface, handles O2O or indoor to indoor transmissions and performs the pathloss computation in the \texttt{GetLoss} method. In this class it is possible to deterministically set the LOS condition, using the \texttt{ChannelCondition} attribute, so that the UT is always in a LOS or NLOS condition, or opt for the statistical approach proposed by the 3GPP model. With the latter, given the selected scenario, the 2D distance and the height of the BS and UT, it is possible to compute a LOS probability $P_{LOS}$ using~\cite[Table~7.4.2-1]{38900}. This is then compared to a random value, $p_{REF}$, generated from a $U(0,1)$ distribution, and the LOS condition for the link is set. In the current implementation of the statistical model, the spatial consistency for different BS-UT links is not modeled, i.e., close-by users have independent LOS conditions. Since this is optional in the 3GPP model, we plan to add it in future releases. The pathloss for each scenario is then computed according to~\cite[Table~7.4.1-1]{38900}. For the UMa, UMi and Indoor-Office scenarios we support both the default and optional NLOS propagation models, and the latter is selected by setting the \texttt{OptionalNlos} attribute to true. Another optional component is the shadowing, which is enabled by the \texttt{Shadowing} attribute. For a moving UT, the shadowing is correlated in space. Given the distance $\Delta d_{2D} > 0$ on the horizontal plane from the last position in which the shadowing was computed, the exponential correlation parameter is computed as
 	$R(\Delta d_{2D}) = e^{-\Delta d_{2D}/d_{cor}},$
with $d_{cor}$ the correlation distance~\cite{recommendation2007prediction}, specified in~\cite[Table~7.5-6]{38900} for each different scenario. Then a first order filter is applied to get the updated shadowing value
\begin{equation}
	s_t = R(\Delta d_{2D})s_{t_{prev}} + \left(\sqrt{1-R^2(\Delta d_{2D})}\right)\sigma_{SF}z, \quad z \sim N(0,1),
\end{equation}
with $s_{t_{prev}}$ the shadowing value in the previous position, and $\sigma_{SF}$ the scenario-dependent shadowing standard deviation. The pathloss and the shadowing (if enabled) are updated at every transmission, while the LOS condition is fixed for the whole duration of the simulation. Fig.~\ref{fig:pathloss} shows the pathloss in dB for the 3D distance from the smallest value supported in each scenario to $10^3$ m for the outdoor ones and $10^2$ m for the indoor ones.

\begin{figure}
\setlength{\belowcaptionskip}{-0.6cm}
	\centering
	\includegraphics[width=0.7\columnwidth,trim = 18cm 2cm 16cm 4cm,clip ]{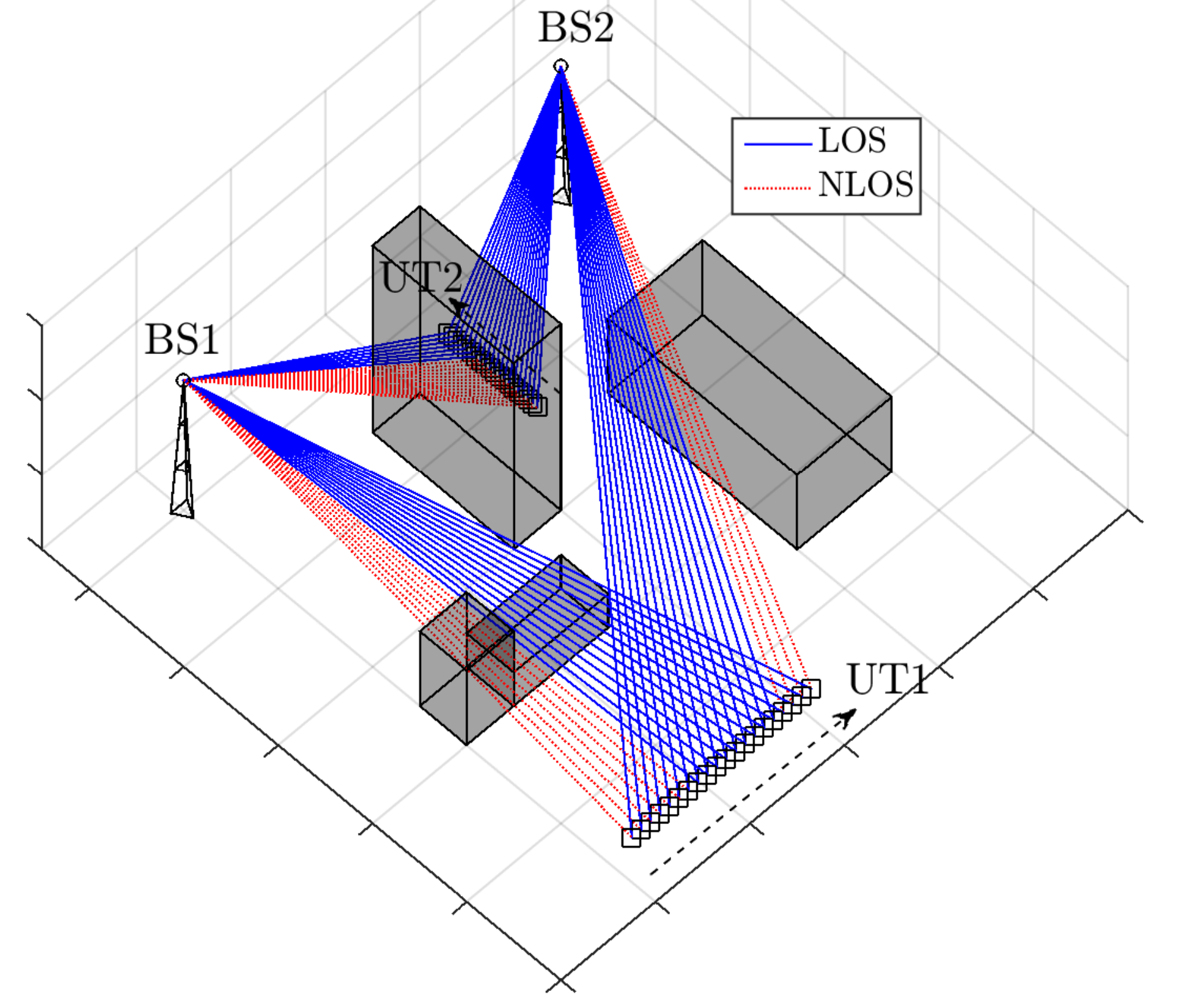}
	\caption{Example of simulation scenario, with 3D buildings that block the LOS between BSs and moving UTs.}
	\label{fig:building}
\end{figure}

The class \texttt{MmWave3gppBuildingPropagationLossModel} provides an alternative model for pathloss computation. This class takes advantage of the ns--3 Buildings module in order to geometrically model the LOS/NLOS and the indoor/outdoor conditions. In particular, it is possible to place buildings in the simulation scenario, randomly or deterministically, and, once the UTs and BSs positions are specified, a 3D ray tracing algorithm determines whether there is LOS or not. An example is shown in Fig.~\ref{fig:building}, where, as UT1 moves, the state of the link to BS1 changes from NLOS (because of the high building that shadows the BS) to LOS. The same happens for UT2. This allows to correlate the LOS/NLOS condition in simulations where the mobility is a key element, and allows to test complex scenarios such as that of Fig.~\ref{fig:building}, or a Manhattan Grid, and so on. Moreover, if the UT is placed inside a building, it is possible to simulate O2I penetration losses. This also implies that BSs must be placed outside of buildings. To simulate indoor to indoor (i.e., the different offices scenarios) communications, the \texttt{MmWave3gppPropagationLossModel} must be used. The 3GPP channel model specifies two different O2I penetration loss models, i.e., low-loss and high-loss. Both are supported by the \texttt{MmWave3gppBuildingPropagationLossModel}, and automatically selected according to the scenario and the kind of building deployed. As specified in~\cite[Sec~7.4.3]{38900}, the low-loss model is used for RMa, while for the UMa and the UMi scenarios the high-loss is selected when the \texttt{BuildingType} is commercial or office, and the low-loss for the residential option. Once the LOS/NLOS condition is set and the possible penetration loss is computed, this class relies on the shadowing and pathloss computation performed in the \texttt{MmWave3gppPropagationLossModel} class. Fig.~\ref{fig:pathloss} also shows an example of the O2I propagation loss for the 3 outdoor scenarios with residential building type.

\subsection{Fast Fading Modeling}
\label{sec:ff}
The fast fading modeling is implemented in the class \texttt{MmWave3gpp\-Channel}, and follows the step by step approach of~\cite[Sec.~7.5]{38900}. In the following we will describe these steps in relation with our implementation, and explain the assumptions and simplifications we introduced in order to reduce the computational load. The fast fading modeling is indeed the bottleneck of system level simulations with wireless channels.

The main method of this class is \texttt{DoCalcRxPower\-Spectral\-Density}, which implements the interface specified in \texttt{SpectrumPropagation\-LossModel} and returns a \texttt{SpectrumValue} object, i.e., a vector with the values of the PSD for each subcarrier of the OFDM symbol transmitted. In order to get this PSD, two steps are required: the first is the actual computation of the channel matrix $\mathbf{H}(t, \tau)$ with the large scale and small scale fading effects, and the second is the generation of the beamforming vectors and of the beamforming gain. This last step will be described in Sec.~\ref{sec:bf}.


Since the channel matrix computation requires several steps and generates several outputs besides the matrix $\mathbf{H}(t, \tau)$, as for example the delay $d_n$ for each cluster, the angles of arrival and departure for each cluster and so on, the auxiliary data structure \texttt{Params3gpp} is used to store this information and possibly re-use it at a later moment (for example, in the spatial consistency procedure). The main parameters of the structure are detailed in Table~\ref{tab:params3gpp}. The output of the channel coefficient generation is the 3D matrix $\mathbf{\hat{H}}$, stored in \texttt{m\_channel}, of size $U\times S\times N$, with the entry $\hat{h}_{u,s,n}$ representing the channel coefficient for cluster $n$ between transmit antenna $s$ and receive antennas $u$. The matrix $\mathbf{H}(t, \tau)$ can then be obtained summing over the $N$ clusters. The Azimuth Angle of Arrival (AoA), of Departure (AoD), Zenith Angle of Arrival (ZoA) and of Departure (ZoD) are stored for cluster $n$ in this order in the $n$-th column of the matrix in \texttt{m\_angle}. Another data structure associated to the \texttt{MmWave3gppChannel} class is a table containing the parameters of~\cite[Table~7.5-6]{38900}, which can be generated at run time for each scenario, given the LOS and indoor/outdoor conditions, the height of the transmitter and receiver and the 2D distance between them.

\begin{table}[!t]
\renewcommand{\arraystretch}{1}
\setlength{\belowcaptionskip}{0.1cm}
\setlength{\abovecaptionskip}{-0.75cm}
\centering
\small
\begin{tabular}{@{}ll@{}}
\toprule
Param & Description \\ \midrule
\texttt{m\_txW} & complex vector with antenna weights for the tx \\
\texttt{m\_rxW} & complex vector with antenna weights for the rx \\
\texttt{m\_channel} & complex 3D matrix $\mathbf{\hat{H}}$  \\
\texttt{m\_delay} & double vector with cluster delays $d_n, n\in[1, N]$  \\
\texttt{m\_angle} & double matrix with AoA, ZoA, AoD and ZoD  \\
\texttt{m\_longTerm} & complex vector with long term fading parameters  \\
\bottomrule
\end{tabular}
\caption{Main entries of the \texttt{Params3gpp} data structure.}
\label{tab:params3gpp}
\end{table}

The channel parameters procedure described in Sec.~\ref{sec:3gppchannel} starts in the \texttt{DoCalcRxPowerSpectralDensity} method. The information on the UT and BS mobility is given as a parameter of this function, thus it is possible to compute the relative 2D and 3D distance, speed, and the horizontal and vertical angles between the two. The scenario, the LOS condition and the possible indoor/outdoor state are obtained from the specific implementation of \texttt{PropagationLossModel} which is used in the simulation and associated to the channel model class, so that the link status is consistent across the different steps of the channel coefficients and pathloss computation. This covers steps 1-3 of the procedure in~\cite[Sec.~7.5]{38900}. Then the actual generation of the matrix is carried out in the \texttt{GetNewChannel} method. Notice that, if the scenario is static, we generate the channel matrix once for each simulation, while if there is mobility the spatial consistency procedure is used, as it will be described in the next section.

The fourth step requires the generation of correlated large scale fading parameters (LSPs), as described in Sec.~\ref{sec:3gppchannel}. In order to speed up the computation, we generated the correlation matrix $\sqrt{\mathbf{C}}$ for each scenario, LOS/NLOS condition and outdoor/indoor state using MATLAB, and saved them in static matrices. The MATLAB code can be found in the ns--3 mmWave module repository. Once the LSPs are computed, it is possible to generate the cluster delays $d_n$ (step 5) and powers (step 6). In case of LOS, the first cluster has a power given by $K_R/(K_R+1)$, with $K_R$ the Ricean factor generated in step 4, and the others are scaled by $1/(K_R+1)$. As specified by the 3GPP model, we remove the clusters whose power is 25 dB smaller with respect to the strongest one, thus, from this step onwards we consider the number of valid clusters $N_v \le N$. In step 7 the AoA, AoD, ZoA and ZoD for each ray in each cluster are generated, scaled in the correct interval ($[0, 360]$ degrees for the horizontal angles and $[0, 180]$ degrees for the vertical one), and finally converted into radians. Then the vectors of angles are randomly shuffled in step 8, in order to introduce a random coupling between AoA and AoD, and ZoA and ZoD. The ninth step is skipped, since in our implementation we consider only antennas with vertical polarization. This impacts also step 10, since we draw only the vertical random initial phase $\Phi^{\theta\theta}_{m,n} \sim U(-\pi,\pi), \; \forall \, m \in [1, M], n \in [1, N_v]$. We also draw an initial random phase for the vertical polarization for the LOS cluster.

Finally, in the eleventh step, for each transmit and receive antenna elements and each cluster an entry of the channel matrix is generated. The factor that accounts for the antenna radiation pattern is obtained for each antenna element from the \texttt{AntennaArray\-Model} object associated to either the transmitter or the receiver. In the generation of the channel matrix entries we introduce a simplification, in order to reduce the computational complexity: the Doppler effect is not computed for each ray, but only for the central angle of each cluster, and in the \texttt{CalBeamformingGain} function, i.e., when the beamforming gain is computed and applied to the PSD. The other parameters are computed for each ray, which are then summed together in order to get the parameter of the whole cluster. We also account for the additional modeling required for the two strongest clusters, which are split into three sub-clusters with different and fixed sub-cluster delays. Finally, the \texttt{GetNewChannel} method stores all the parameters and returns a \texttt{Params3gpp} structure.

\section{Beamforming Scheme}
\label{sec:bf}
In this section we describe the analog MIMO beamforming scheme implementation that can be used together with the channel model described in Sec.~\ref{sec:channelImpl}. Once pathloss, shadowing and channel coefficient matrix are generated, the beamforming vector at transmitter and receiver and the gain are computed and the latter is applied to the PSD object handled by the \texttt{DoCalcRxPower\-Spectral\-Density} method of \texttt{MmWave3gppChannel}. Our framework assumes analog beamforming and a TDD physical layer frame, as described in~\cite{ford2016framework}. Since with analog beamforming a single stream is transmitted, each antenna element only applies a phase shift to the transmit signal, so that it is concentrated in one direction. The beamforming gain is obtained by multiplying the transmitter and receiver beamforming vectors with the channel coefficients matrix $\mathbf{H}(t,\tau)$. 

In the current implementation we provide two ways to compute the beamforming vector, but we designed the framework so that researchers that would like to test different beamforming solutions can easily extend the \texttt{MmWave3gppChannel} with their own solutions. The two default methods will be briefly described in the following paragraphs, and it is possible to select one of the two using the \texttt{CellScan} attribute.

\subsection{Beamforming Vector Calculation}

\textbf{Power Method:} the power method, which is enabled by default, computes the optimal beamforming vector for the transmitter and the receiver, assuming a perfect knowledge of the channel matrix, and it is implemented in the \texttt{PowerMethodBeamforming} method. It is based on the MIMO Maximum Ratio Transmission scheme, for which the transmitter and receiver BF vectors are the eigenvectors associated to the largest eigenvalues of spatial correlation matrix $\mathbf{Q}_{tx}$ and $\mathbf{Q}_{rx}$, respectively~\cite{love2003equal}. Therefore, for the transmitter we first sum over the valid clusters the matrix $\mathbf{\hat{H}}$ of the \texttt{Params3gpp} structure, in order to get a two-dimensional matrix $\mathbf{H}$ of size $U\times S$, and then compute $\mathbf{Q}_{tx} = \mathbb{E}[\mathbf{H}^{\dagger}\mathbf{H}]$, where $^{\dagger}$ is the conjugate transpose operation and the expectation is taken with respect to the small scale fading. Then, in order to identify the correct eigenvector, we use the power method~\cite{wilkinson1965algebraic}. The algorithm selects a random initial beamforming vector and iteratively multiplies it with the spatial correlation matrix $\mathbf{Q}_{tx}$, normalizing the results at each iteration. Finally, the output will converge to the correct eigenvector. As a stopping criterion, we check the difference between successive vectors against a threshold, and set a maximum number of iterations $R$. The computation for the receiver is done in the same way, starting from $\mathbf{Q}_{rx} = \mathbb{E}[\mathbf{H}\mathbf{H}^{\dagger}]$. In our simulator, we always assume that the transceiver knowns the channel matrix and is able to update the beamforming vector as soon as the channel changes. However, the developer may extend the \texttt{PowerMethodBeamforming} method in order to control the beamforming vector update frequency, or even introduce a delay between the updates of the channel and those of the beamforming vectors.

\begin{figure}[t]
	\centering
	\includegraphics[width=0.9\columnwidth,trim = 0.5cm 1.5cm 1.5cm 0cm,clip ]{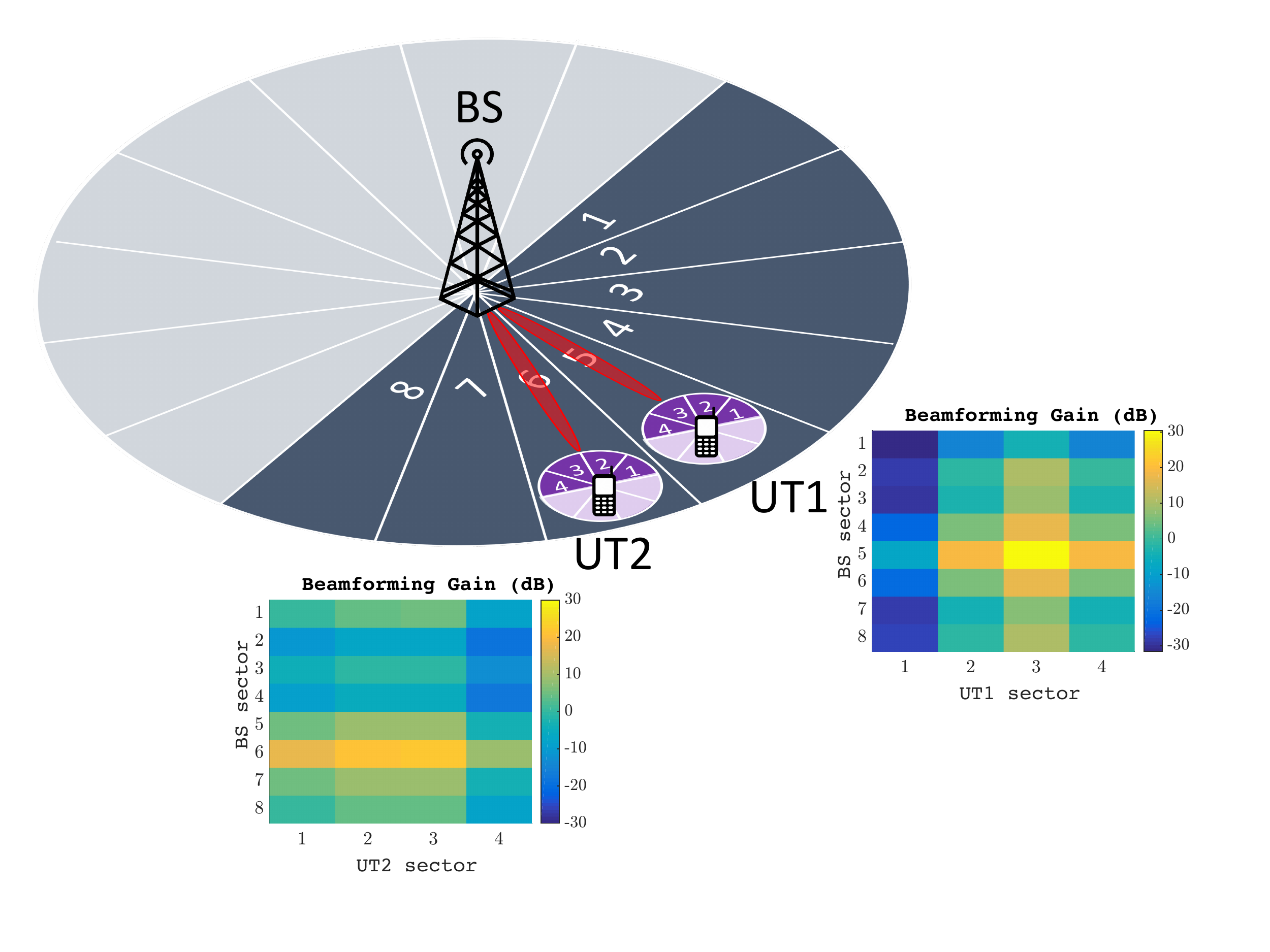}
	\caption{Example of sector selection with cell scanning method.}
	\label{fig:cellscan}
\end{figure}

\textbf{Cell Scanning Method:} the cell scanning method does not assume that the transmitter and receiver known the channel information. A variant of this solution is implemented in the IEEE 802.11ad standard for the 60 GHz band~\cite{ieeeWifi}. Basically, the cell scanning method divides the coverage area of the cell into several sectors and steers the beam to one of the directions by selecting the beamforming vector from a predefine codebook. This means that, if the transmitter wants to transmit in sector $M$, it selects from a list (i.e., the codebook) the beamforming vector corresponding to that sector and the signal will be directed in that sector. The implementation of the UPA antenna, the \texttt{AntennaArrayModel} class, supports any arbitrary number of antenna elements, and the beamforming vector for each sector is generated by calling the \texttt{SetSector} method.

Fig.~\ref{fig:cellscan} shows an example of one BS with 64 antenna elements and two UTs with 16 antenna elements. The maximum beamforming gain associated to this setup is around 30 dB (i.e., roughly equal to the product of transmit and receive antennas). The scheme shown in Fig.~\ref{fig:cellscan} assumes that all the BSs and UTs define as sector $\xi=1$ the one pointing to northeast, and that the number of sectors for an antenna panel is equal to the number of horizontal antenna elements (i.e., 8 and 4). LOS condition is maintained during the whole simulation. If we consider the BS-UT1 link, then the LOS cluster angles of arrival and departure are such that the its rays point to the center of a sector both in the UT and in the BS, thus the maximum gain of 30 dB can be achieved by selecting BS sector $\xi=5$ and UT sector $\xi=3$. If this does not happen, the gain is reduced. For example, if we consider the link between the BS and UT2, the LOS cluster points to the center of BS sector $\xi=6$, but on the UT side it is directed on the edge of sectors $\xi=2$ and $\xi=3$, therefore the maximum gain that could be achieved is 20 dB, when either UT sector $\xi=2$ or $\xi=3$ is selected.

The method \texttt{CellScan} of the class \texttt{MmWave3gppChannel} currently implements a simplified cell scanning approach, based on a brute force approach, i.e., it tries all the possible BS-UT sector combinations and selects the one that returns the maximum received power. \texttt{CellScan} only shows how our MIMO architecture can be used to develop and test cell scanning algorithms, in order to encourage researchers and developers interested in the beam-tracking problem to implement and test more refined solutions in the ns--3 mmWave module.

\subsection{Beamforming Gain}
The channel matrix is generated for all the possible BS-UT links, but the beamforming vectors are computed only for the links involved in a communication, i.e., those for which a UT is attached to a BS. Therefore, the antenna beams are perfectly aligned for the communication links, but not for interference links. After the beamforming vector is selected for each device, the final gain $G(t,f_s)$ at time $t$ and subcarrier frequency $f_s$ is computed as
\begin{equation}
	G(t,f_s) =\sum_{n=1}^{N}{\bf w}_{rx}^{\dagger}{\bf H}_n {\bf w}_{tx}e^{j2\pi \frac{\mathbf{r}_{rx,n}^{T}\mathbf{v}t}{\lambda_0}}e^{j2\pi d_nf_s},
\end{equation} 
where ${\bf w}_{rx}$ and ${\bf w}_{tx}$ are the receive and transmitter side beamforming vectors, ${\bf H}_n$ is the matrix of size $U\times S$ representing the long term MIMO channel for cluster $n$ (i.e., a slice of matrix $\mathbf{\hat{H}}$ computed in Sec.~\ref{sec:ff}). The first exponential term is the small scale fading effect caused by the Doppler phenomenon, which is applied only to the central angle of the cluster, as mentioned in Sec.~\ref{sec:ff}. $\mathbf{r}_{rx,n}^{T}$ is the transpose of the receiver spherical unit vector, with azimuth arrival angle $\phi_{n,AOA}$ and elevation arrival angle $\theta_{n,ZOA}$, given by
\begin{equation}
r_{rx,n}=\begin{bmatrix}
\sin{\theta_{n,ZOA}}\cos{\phi_{n,AOA}}\\ 
\sin{\theta_{n,ZOA}}\sin{\phi_{n,AOA}}\\
\cos{\theta_{n,ZOA}},
\end{bmatrix},\end{equation}
$\mathbf{v}$ is the vector of the relative speed in a 3D space between the UT and the BS and $\lambda_0$ is the wavelength. The second exponential term represents the frequency selectivity effect caused by the delay $d_n$ of the $n$-th cluster.

Finally, notice that the beamforming gain $G(t,f_s)$ may be rewritten as 
\begin{equation}
	G(t,f_s) =\sum_{n=1}^{N}L_n \times F_n(t,f_s),
\end{equation} 
where $L_n = {\bf w}_{rx}^{\dagger}{\bf H}_n {\bf w}_{tx}$ represents the long term components of the fading, and $F_n(t,f_s) = e^{j2\pi \mathbf{r}_{rx,n}^{T}\mathbf{v}t/\lambda_0}e^{j2\pi d_nf_s}$ is the only component that depends on the subcarrier frequency and time. Therefore, the beamforming gain computation can be optimized by storing $L_n$ for each cluster, and updating only $F_n(t,f_s)$ for each transmission. However, notice that the term $L_n$ is recomputed every time the channel or the beamforming vectors are changed.

\section{ns--3 Implementation of Optional Features}
\label{sec:opt}

\subsection{Spatial Consistency}

The basic channel model described in the previous sections can be used for drop-based simulations with limited mobility, i.e., for UTs that move in an area in which the channel is very correlated and the fading parameters do not change. However, for simulations in which the mobility is an important factor, the spatial consistency of the channel throughout the path on which the UT moves can be simulated by enabling this option in the \texttt{MmWave3gppChannel} class. In the current implementation, we support spatial consistency with Procedure A of~\cite[Sec~7.6.3.2]{38900} for both LOS and NLOS communications. The extension with the soft LOS state for the LOS/NLOS transition will be added soon.

The spatial consistency procedure is implemented in the \texttt{Update\-Channel} method. It is possible to set the period of update $t_{PER}$ by changing the \texttt{UpdatePeriod} attribute. Then, every $t_{PER}$, the particular realization of the channel matrix is deleted, and the \texttt{DoCalcRx\-PowerSpectralDensity} method is forced to recompute it by calling the \texttt{UpdateChannel} method. The latter updates the cluster delays, powers and departure and arrival angles computed at the previous update with a transformation that accounts for the speed of the UT and for the distance traveled on the horizontal plane. Notice that, according to the procedure described in~\cite{38900}, the channel should be updated taking into account the one generated at the beginning of the simulation, i.e., when the UT is dropped in the scenario. However, the UT may change its velocity and trajectory during the ns--3 simulation, and an update that considers only the initial settings may be inaccurate. Therefore the channel at time $t$ is updated with respect to that generated at time $t-t_{PER}$.

\begin{figure}[!t]
\setlength{\belowcaptionskip}{-0.7cm}
	\centering
	\includegraphics[width=0.9\columnwidth,trim = 22cm 1.1cm 1cm 0cm,clip ]{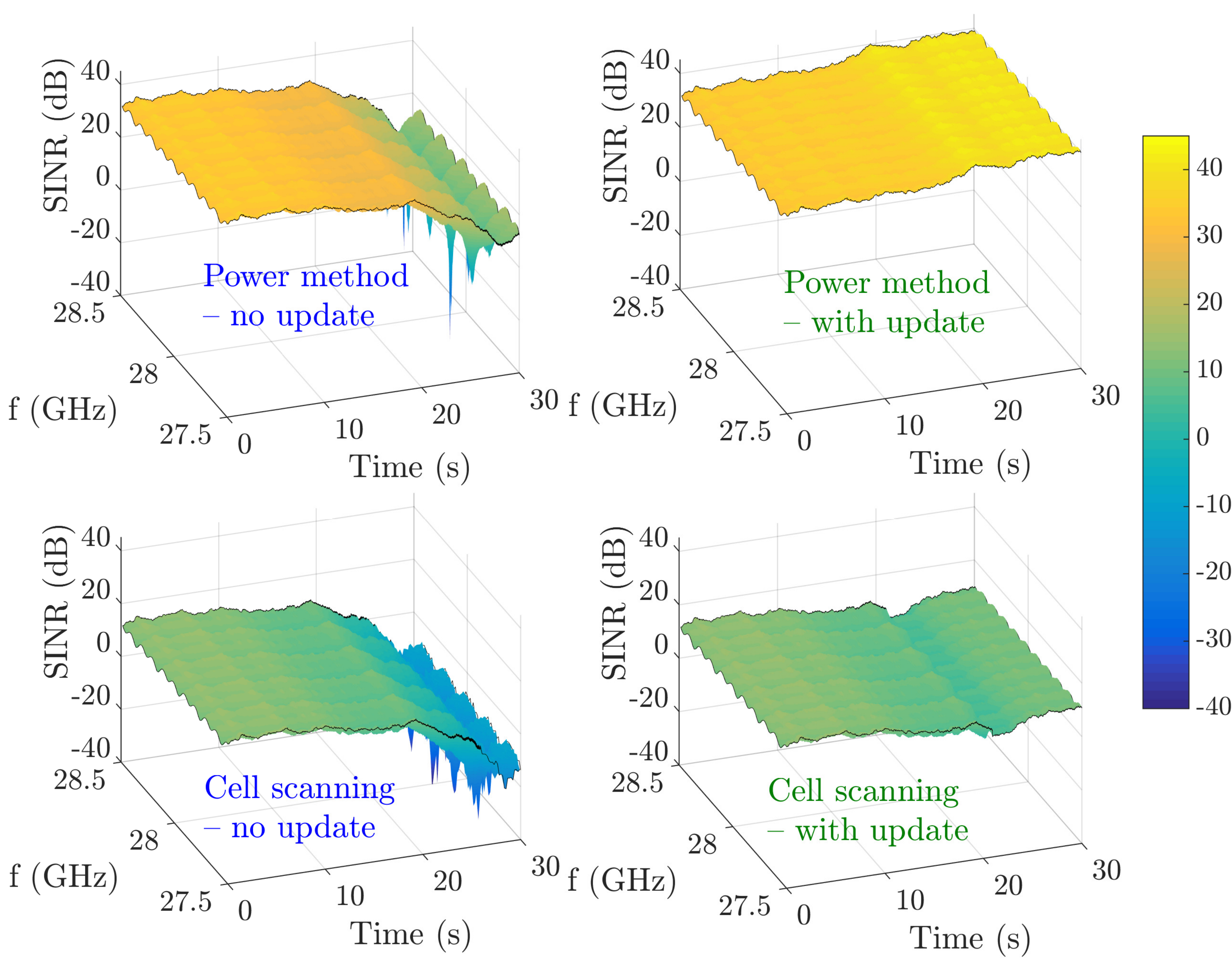}
	\caption{Moving UT SINR over frequency and time with power method and cell scanning method,  with and without updating.}
	\label{fig:beamforming}
\end{figure}

Fig.~\ref{fig:beamforming} shows an example of a RMa scenario with a BS at coordinates $(0,0,35)$~m and a UT in position $(100, 0, 1.5)$~m, and moving at 1 m/s along the y axis and maintaining LOS connectivity. The channel is updated with consistently every $t_{PER}=100$~ms. The 2 figures in the left column shows the SINR that is obtained without updating the BF vectors computed at $t=0$~s, and this causes the SINR to drop as UT moves. Notice that the current implementation of the cell scan method uses a fixed elevation elevation angle of 90 degrees and swipes only the horizontal plane, and this is why it has 20 dB less power. The two plots on the right instead show the SINR when the BF vector is updated together with the channel. It is possible to observe that for the first 20 s the performance with and without the BF update is similar, because of the consistency of the channel and of the low mobility of the UT, but after $t = 20$ s the SINR without update degrades by nearly 20 dB. Moreover, the power method finds the optimal BF vector whenever the channel is updated, therefore the SINR is very stable. Even though, the UT goes further away from BS over time, the SINR actually increased 10 dB due to shadowing. On the other hand, the cell scanning method shows a loss in SINR, because when the UT moves it cannot optimally adapt the BF vector but just select one of the available sectors.

\subsection{Blockage}

This optional feature can be used to model the attenuation in certain clusters, according to their angle of arrival. The attenuation can be caused by the human body that holds the UT, or by external elements as for example cars, other human bodies, trees. The blockage model is implemented in the \texttt{ApplyBlockageModel} of the \texttt{MmWave3gppChannel} and activated using the \texttt{Blockage} attribute. In our implementation we consider blockage model A, which only distinguishes between self-blocking and non-self-blocking, and is generic and computationally efficient~\cite{38900}. In particular, this model randomly generates $K+1$ blocking regions, one for the self-blocking, with different parameters according to the orientation of the UT (i.e., portrait or landscape mode, customizable with the \texttt{PortraitMode} attribute), and $K$ for the non-self-blocking (with $K=4$ as default, controlled by the \texttt{NumNonSelfBlocking} attribute). The attenuation is of 30 dB for self-blocking, and dependent on the scenario, on horizontal and vertical arrival angles for non-self-blocking. Moreover, the blocking of a certain cluster is correlated in both space and time, according to the UT mobility and the scenario of the simulation. Notice that, if both the blockage and the spatial consistency options are used, then the update of the channel with both features is synchronized, i.e., the cluster blockage is updated before the channel coefficients are recomputed with the spatial consistency procedure.

\begin{figure}[t]
\setlength{\belowcaptionskip}{-0.4cm}

	\centering
	\includegraphics[width=0.9\columnwidth,trim = 1cm 1.5cm 1cm 1.5cm,clip ]{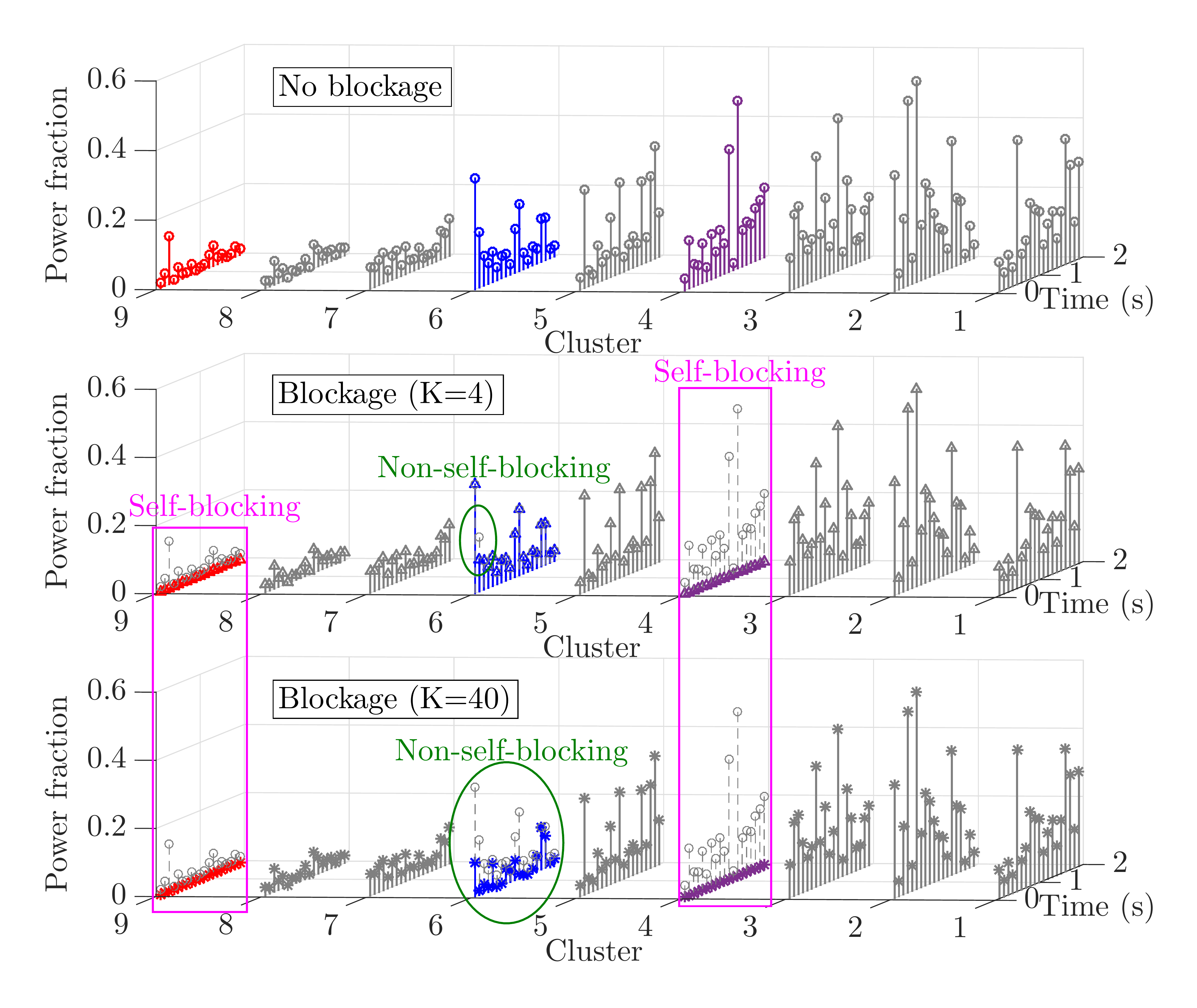}
	\caption{Cluster power fraction with and without blockage}
	\label{fig:blockage}
\end{figure}

An example of results can be seen in Fig. \ref{fig:blockage}, which shows the evolution of the power of each cluster in time for a different number of blockers $K+1$. Spatial consistency is also applied, with an update every $t_{PER}=100$ ms, as shown in the top plot (i.e., the one without blockage). When blockage is introduced, and $K=4$, clusters 4 and 9 are blocked by self-blocking and only a single realization in time of cluster 6 is blocked by non-self-blocking. It is clear that self-blocking has a larger blocking region and higher attenuation, therefore it has a stronger impact on the cluster power. Moreover, when $K=40$, cluster 6 suffers more from the non-self-blocking phenomenon. Notice also that the other clusters are not affected by blockage, even with the increased non-self-blocking region, because the elevation of each non-self-blocking region is relatively narrow (i.e., max 10 degrees for outdoor scenarios and 30 degrees for indoor). 

\section{Conclusions}
\label{sec:conclusions}
In this paper, we described the implementation of the 3GPP channel model for the 6-100 GHz band in the mmWave ns--3 framework. According to 3GPP, this channel model is the one that should be used for system level simulations in order to assess the performance of proposed 5G NR technologies. Moreover, it is an improvement over the previous models that the framework offered, since it covers a wider range of frequencies, it allows to model the channel consistency for simulations with mobility along with random blockers that attenuate the received signal. We provide the source code and a tutorial example in the mmWave ns--3 repository. As future work, we plan to extend our implementation in order to enable the integration of these models with WiFi and other wireless modules. Besides, we plan to implement other optional features in order to make the channel model more flexible and suitable for a wide range of simulations that involve the 6-100 GHz band.

\bibliographystyle{ACM-Reference-Format.bst}
\bibliography{bibl}

\end{document}